\documentclass[prd,twocolumn,showpacs,preprintnumbers,floats,
amssymb,amsmath,nofootinbib]{revtex4}
\usepackage{graphicx}
\usepackage{dcolumn}
\usepackage{bm}
\def\spose#1{\hbox to 0pt{#1\hss}}
\def\lta{\mathrel{\spose{\lower 3pt\hbox{$\mathchar"218$}}
     \raise 2.0pt\hbox{$\mathchar"13C$}}}
\def\gta{\mathrel{\spose{\lower 3pt\hbox{$\mathchar"218$}}
     \raise 2.0pt\hbox{$\mathchar"13E$}}}
\newcommand{\be}{\begin{equation}}
\newcommand{\en}{\end{equation}}
\newcommand{\bea}{\begin{eqnarray}}
\newcommand{\ena}{\end{eqnarray}}

\begin{document}
\draft

\title{The Corley-Jacobson dispersion relation and trans-Planckian 
inflation}

\author{J\'er\^ome Martin} 
\email{jmartin@iap.fr}
\affiliation{Institut d'Astrophysique de Paris, 98bis 
boulevard Arago, 75014 Paris, France}
\author{Robert H. Brandenberger}
\email{rhb@het.brown.edu}
\affiliation{Theory Division, CERN, CH-1211 Geneva 23, Switzerland\footnote{on leave 
from absence: Department of Physics, Brown University, Providence, 
RI 02912, USA}}
\date{\today}

\begin{abstract} 
In this Letter we study the dependence of the spectrum of fluctuations
in inflationary cosmology on possible effects of trans-Planckian
physics, using the Corley/Jacobson dispersion relations as an
example. We compare the methods used in previous work
\cite{Martin:2001xs} with the WKB approximation, give a new exact
analytical result, and study the dependence of the spectrum obtained
using the approximate method of \cite{Martin:2001xs} on the choice of
the matching time between different time intervals. We also comment on
recent work subsequent to \cite{Martin:2001xs} on the trans-Planckian
problem for inflationary cosmology.
\end{abstract}
\maketitle

\section{Introduction}

The trans-Planckian problem of inflation is the following: in many
models of inflation, the phase of accelerated expansion lasts so many
$e$-foldings that the comoving lengths corresponding to present day
cosmological scales were much smaller than the Planck length at the
beginning of inflation. Hence, one may wonder whether the ``standard"
predictions of inflation, in particular the fact that the power
spectrum of cosmological perturbations is close to scale-invariant,
will be changed if the laws of physics beyond the Planck scale are
different from the ones which rule the low energy phenomena. The usual
calculations of the spectrum of cosmological perturbations (see
e.g. \cite{Mukhanov:1992me} for a comprehensive review) is based on
the use of classical general relativity coupled to a weakly
interacting scalar field, and on linearizing the resulting equations
of motion about a classical background cosmology. The validity of this
approach in the trans-Planckian regime is highly doubtful.
\par

In the context of inflation, this question was first addressed in
Ref.~\cite{Martin:2001xs}.  To calculate the power spectrum of
cosmological perturbations, the main equation that needs to be solved
is the equation of a parametric oscillator with a time-dependent
frequency which is a function of the scale factor $a(\eta )$ (and its
derivatives) and of the dispersion relation $\omega _{_{\rm
phys}}(k)$, where $k$ indicates the physical wave number related to
the comoving wave number $n$ by $k=n/a$. In the usual discussions, the
dispersion relation is taken to be linear as appropriate for a free
field theory. The method used in Ref.~\cite{Martin:2001xs} was to
replace the linear relation $\omega _{_{\rm phys}}= k$ by a
non-standard one that mimics possible modifications of the physics in
the ultraviolet regime. Based on similar approaches
\cite{U,Corley:1996ar} used to study the possible dependence of
Hawking radiation on trans-Planckian physics, two classes of
dispersion relations were considered [$n_{\rm eff}\equiv a(\eta
)\omega _{_{\rm phys}}(n/a)$]:
\begin{eqnarray}
\label{disp}
n_{\rm eff} &=& n\frac{\lambda }{\ell _{_{\rm C}}}\tanh
^{1/p}\biggl[\biggl( \frac{\ell _{_{\rm C}}}{\lambda
}\biggr)^p\biggr], 
\\ 
n_{\rm eff} &\equiv & \sqrt{n^2+n^2b_1
\biggl(\frac{\ell _{_{\rm C}}}{\lambda }\biggr)^{2}},
\end{eqnarray}
where $\lambda $ denotes the physical wavelength of a given mode, and
$\ell _{_{\rm C}}$ is a characteristic length expected to be
determined by the Planck scale. The first one is the Unruh dispersion
relation \cite{U} whereas the second one is the Corley/Jacobson
relation \cite{Corley:1996ar} ($b_1$ is an arbitrary number which can
be positive or negative). In fact, in Ref.~\cite{Martin:2001xs} a
generalization of the Corley/Jacobson dispersion relation was
considered.
\par

In \cite{Martin:2001xs}, the problem was investigated for the class of
scale factors corresponding to power-law inflation, i.e. $a(\eta
)=\ell _0\vert \eta \vert ^{1+\beta }$, $\beta \le -2$ where $\ell _0
$ has the dimension of a length and is equal to the Hubble radius
during inflation if $\beta =-2$ (de Sitter inflation). The example of
the Unruh dispersion relation was treated only for $\beta =-2$ whereas
the second dispersion relation was studied for any value of $\beta \le
-2$. It was found that no modifications in the spectrum of
fluctuations arise in the first case, whereas some differences can
show up in the second case if $b_1<0$\footnote{This result was
obtained assuming that the initial state is the ``minimizing energy
state''. In Ref.~\cite{Martin:2001xs}, another state was also
considered, but only to demonstrate that the final spectrum depends on
the choice of the initial state. As stressed in
Ref.~\cite{Martin:2001xs}, the ``minimizing energy state'' is the only
physically well-motivated state.}.
\par 

The aim of this letter is to return to the example of the
Corley/Jacobson dispersion relations, making use of a new exact
analytical solution. This one allows us to address some technical
points that have been raised recently in the literature and to compare
the method of Ref.~\cite{Martin:2001xs} with the other methods used in
the works subsequent to \cite{Martin:2001xs}. Explicitly, in Section
II, we compare the Wigner-Kramers-Brillouin (WKB) approximation method
used in Ref.~\cite{Niemeyer:2001qe} to calculate the power spectrum
with the method used in Ref.~\cite{Martin:2001xs}. In section III, we
consider the exact formula mentioned above which allows us to make a
smooth transition from the region where the sub-Planckian effects are
important to the region where the dispersion relation is standard. We
take advantage of the fact that this new solution is also valid for
$b_1<0$ to discuss in more detail than in Ref.~\cite{Martin:2001xs}
the physical meaning of complex solutions in the sub-Planckian
region. In the fourth section, we study the consequences for the
matching time used in the approximate analysis of \cite{Martin:2001xs}
and show that the use of an incorrect matching time will lead to
artificial oscillations in the spectrum\footnote{As pointed out in
\cite{Niemeyer:2001qe}, there was a mistake in Section V.B.2 of the
first version of \cite{Martin:2001xs}: the incorrect choice of the
matching time in the Corley/Jacobson case with $b_1>0$ led to a
spectrum which was the usual one times a complicated oscillatory
function, instead of to an unmodified spectrum.}. First, however, we
comment on some of the recent work on this subject.
\par

Following Ref.~\cite{Martin:2001xs}, there has been a significant
amount of work on the trans-Planckian problem for inflationary
cosmology.  The work has focused on two issues. The first is how broad
the class of dispersion relations is for which there is a change in
the spectrum of fluctuations, and whether there are other features in
the linear perturbations which can be used to probe trans-Planckian
physics. The second is whether the back-reaction of the excess
fluctuations produced by modified dispersion relations are important
and can have interesting consequences.
\par

Let us for the moment focus on the issue of the class of dispersion
relations for which a modification of the spectrum of fluctuations is
found. In \cite{Niemeyer:2001eh}, the case of Unruh dispersion
relations was studied in a de Sitter Universe, and no change in the
spectrum was found, in agreement with the corresponding results in
\cite{Martin:2001xs}. No dispersion relations of Corley/Jacobson type
were studied in \cite{Niemeyer:2001eh}\footnote{On this basis, the
claims of Ref. \cite{Niemeyer:2001qe} that ``contradictory results''
or ``opposite results'' were found in Refs.~\cite{Martin:2001xs} and
\cite{Niemeyer:2001eh} are misleading. Indeed, the only common case
between these two articles is the case of the Unruh dispersion
relation in de Sitter spacetime for which exactly the same conclusion
was obtained in Refs.~\cite{Martin:2001xs} and
\cite{Niemeyer:2001eh}.}. In \cite{Niemeyer:2001qe,Starobinsky:2001kn}
the trans-Planckian problem of inflationary cosmology was addressed
assuming that the mode wave function is of WKB-type form. Once again,
no changes in the spectrum were found. It has been pointed out in
\cite{Martin:2000bv} that for dispersion relations which lead to
adiabatic evolution of the states on sub-Planckian scales - and WKB
states fall into this category - there are no changes in the spectrum
compared to what is obtained with a linear dispersion relation. As
shown in \cite{Martin:2000bv}, the evolution in the case of the
Corley-Jacobson dispersion relations with a negative sign of $b_1$ is
not adiabatic, and this is the reason that the spectrum is
modified. Note that the change in the spectrum of fluctuations in the
case of Corley/Jacobson dispersion relations with $b_1 < 0$ was
confirmed in \cite{Starobinsky:2001kn}, where these dispersion
relations were labeled as ``exceptional forms of $\omega (k)$''. Other
dispersion relations were studied in
\cite{Kowalski-Glikman:2001dz,Mersini:2001su,Alexander:2001ck,Alexander:2001dr}
\footnote{The goal of \cite{Mersini:2001su} was to provide a new model for
dark energy (see, however, \cite{Lemoine:2001ar}), and in \cite{Alexander:2001ck}
and \cite{Alexander:2001dr} novel dispersion relations were used to yield
realizations of the varying speed of light scenario \cite{Alexander:2001ck}
and to obtain inflation from radiation \cite{Alexander:2001dr}.}. 
\par

The question of a possible dependence of the predictions for linear
cosmological fluctuations on trans-Planckian physics was recently
considered in the context of possible approaches to quantum gravity in
~\cite{Kempf:2001ac,Chu:2000ww,Kempf:2001fa,Easther:2001fi}.  In
\cite{Kempf:2001ac,Kempf:2001fa,Easther:2001fi} the starting point was
modifications of the commutation relations stemming from general
considerations of short-distance quantum gravitational effects, and in
\cite{Chu:2000ww} consequences of short-distance non-commutative
geometry was investigated. Interesting deviations of the spectrum of
fluctuations from the usual results were found. Specifically, the
analysis of \cite{Chu:2000ww} showed that non-Gaussian fluctuations
are expected, and the analysis of \cite{Easther:2001fi} revealed
changes to the spectral shape.
\par

To summarize the current state of knowledge on the issue of the
possible dependence of the spectrum of fluctuations on trans-Planckian
physics, if the dispersion relation leads to adiabatic evolution of
the vacuum state on sub-Planckian length scales, then the spectrum is
not modified \cite{Martin:2000bv,Niemeyer:2001qe,Starobinsky:2001kn}.
However, this restriction on the class of dispersion relations may
exclude the cases of actual physical interest
\cite{Chu:2000ww,Easther:2001fi}.  String theory, M-theory,
non-commutative geometry and discrete quantum gravity can all lead to
much more drastic changes in the effective dispersion relation and may
hence result in changes in the spectrum of fluctuations which can be
probed observationally with current and future CMB experiments. Recently, 
two concrete examples where a change in the spectrum is obtained were 
studied in Refs.~\cite{Lemoine:2001ar,BJM}.
\par

The second issue raised in recent work, in particular in
\cite{Tanaka:2000jw,Starobinsky:2001kn} is whether, 
if the power spectrum is indeed modified
by trans-Planckian physics, there is a back-reaction problem 
\footnote{Clearly, the back-reaction problem can only be raised as a
second question once it has been established that there are important
modifications to the fluctuation spectrum.}. The back-reaction problem
is the following: when viewed at late times, the modified dispersion
relations studied in~\cite{Martin:2001xs} and in subsequent papers
lead to mode functions which during inflation on length scales larger
than $\ell_{_{\rm C}}$ but smaller than the Hubble radius are excited
compared to the adiabatic vacuum. These fluctuations carry energy and
momentum, and this energy could be so large as to turn off inflation,
in a similar manner as the back-reaction of cosmological fluctuations
in models of chaotic inflation~\cite{ABM} can build up and terminate
inflation. This is a very interesting issue which merits detailed
study. 

\section{WKB approximation vs. piecewise solutions}

The equation of motion that needs to be solved in order to compute 
the power spectrum of cosmological perturbations is 
\begin{equation}
\label{paraeq}
\mu ''+\omega ^2(n,\eta )\mu =0 \, ,
\end{equation}
where the expression for $\omega$ is given below [see
Eq.~(\ref{omega})]. Usually, Eq.~(\ref{paraeq}) possesses two regimes
depending on whether the wavelength $\lambda $ of the mode is smaller
or larger than the Hubble radius $\ell_{_{\rm H}}$. In the present
context, we deal with three different regimes. The first one is when
$\lambda \ll \ell _{_{\rm C}} \ll \ell_{_{\rm H}}$. This defines
Region I of Ref.~\cite{Martin:2001xs}. The second regime is such that
$\ell _{_{\rm C}} \ll \lambda \ll \ell_{_{\rm H}}$ and corresponds to
Region II of Ref.~\cite{Martin:2001xs}. Finally, the third regime is
when $\ell _{_{\rm C}} \ll \ell_{_{\rm H}}\ll \lambda$: this is Region
III of Ref.~\cite{Martin:2001xs}. We recover the usual result for the
spectrum of fluctuations if in Region II $\mu$ is given by a single
branch with a coefficient proportional to $1/\sqrt{n}$ multiplying the
plane wave solution.
\par

The method used in Ref.~\cite{Martin:2001xs} to determine the solution
of Eq.~(\ref{paraeq}) is to find solutions in the three regions and to
match them at the boundaries between the regions.  The method used in
Ref.~\cite{Niemeyer:2001qe} is different and consists in utilizing the
WKB approximation. In that case, the solution of Eq.~(\ref{paraeq}) is
given by
\begin{equation}
\label{wkb}
\mu _{\rm wkb}= \frac{1}{\sqrt{2\omega (\eta )}}
\exp \biggl[\pm i\int _{\eta _{\rm i}}^{\eta }
\omega (\tau ){\rm d}\tau \biggr] ,
\end{equation}
where $\eta_{\rm i}$ is some initial time.
\par

In cases where the adiabatic approximation is justified, the two
methods are equivalent. To verify this, consider first Unruh's
dispersion relation. Inserting the dispersion relation into
Eq.~(\ref{wkb}) (using the minus sign) yields in Region I
\begin{equation}
\label{unruhsol}
\mu _{\rm wkb}=\frac{1}{2}\sqrt{\frac{\epsilon }{\pi }}
\vert \eta _{\rm i}\vert ^{1/2}\biggl( \biggl\vert \frac{\eta }{\eta _{\rm i}}
\biggl \vert \biggl)^{-2\pi i/\epsilon +1/2}.
\end{equation}
This is nothing but Eq.~(66) of Ref.~\cite{Martin:2001xs}
[i.e. Eq.~(54) in the limit when $\epsilon \equiv \ell _{_{\rm
C}}/\ell _0$ is small], properly normalized according to
Eqs.~(56)-(57) of that reference. Similarly, in the case of the
Corley/Jacobson dispersion relation (with $b_1>0$, in which case the
adiabatic condition is satisfied~\cite{Niemeyer:2001qe}),
Eq.~(\ref{wkb}) becomes
\begin{widetext}
\begin{equation}
\label{JCwkb}
\mu _{\rm wkb}=2^{-1/2}\biggl( \frac{n\epsilon }{2\pi }\biggr)^{-1/2}
n^{-1/2}b_1^{-1/4}\vert \eta \vert ^{-1/2} \exp 
\biggl[\mp ib_1^{1/2}\biggl(\frac{n\epsilon }{4\pi }\biggr)
n\biggl(\vert \eta \vert ^2
-\vert \eta _{\rm i}\vert ^2\biggr)\biggr].
\end{equation}
\end{widetext}
Again, this is the same result as obtained with the matching technique
in Ref.~\cite{Martin:2001xs}. To see this, take Eq.~(118) of that
reference, which provides the solution in Region I, consider the limit
when $z$ is large and use Eqs.~(126) and (127) with the lower sign. In
Region II, the solution is simply given by plane waves $\mu \simeq
B_1e^{in\eta }+B_2e^{-in\eta }$ in agreement with Eq.~(14) of
Ref.~\cite{Martin:2001xs}.  
\par 

The advantage of the WKB method over the method used in
Ref.~\cite{Martin:2001xs} is that one does not have to perform the
matching between the different regions. The disadvantage is that the
WKB method applies only to examples in which the adiabaticity
condition is satisfied. In the next section, we consider an exact
model which allows us to avoid the matching between Region I and
II. In the third section, we study the consequences for the matching
time and show that an incorrect matching time could lead to artificial
oscillations in the spectrum, as pointed out in
Ref.~\cite{Niemeyer:2001qe}.

\section{The exact model}

The effective time dependent frequency is given by the equation
\begin{equation}
\label{omega}
\omega ^2(n,\eta )=n_{\rm eff}^2-\frac{a''}{a}.
\end{equation}
We consider the Corley-Jacobson dispersion relation given in
Eq.~(\ref{disp}) where, as already mentioned above, $b_1$ is an
arbitrary number which can be positive or negative. We restrict our
consideration to the prototypical model of inflation, i.e. de Sitter
inflation. Then the scale factor can be written as $a(\eta )=\ell
_0/\vert \eta \vert $. If we consider the evolution of the mode well
inside horizon, the term $a''/a$ can be neglected in
Eq.~(\ref{paraeq}). Then, this equation takes on the form
\begin{equation}
\label{mujc}
\mu ''+\biggl[n^2+b_1\biggl(\frac{n^2\epsilon }{2\pi }
\biggr)^2\vert \eta \vert ^2\biggr]\mu =0,
\end{equation}
where $\epsilon \equiv \ell _{_{\rm C}}/\ell _0$. Typically, $\epsilon
$ is a very small number of the order $10^{-5}$. We now need to
distinguish between the cases of positive or negative $b_1$. Let us
first concentrate on the $b_1>0$ case. We make the following change of
variables $-\eta =Cx$ ($x>0$), where $C$ is a constant given by
$C=[(b_1)^{-1/4}/n]\sqrt{\pi /\epsilon}$. Then the equation of motion
takes the form
\begin{equation}
\label{exact}
\frac{{\rm d}^2\mu }{{\rm d}x^2}+\biggl(\frac{x^2}{4}-a\biggr)\mu =0, \quad 
a\equiv -\frac{\pi }{\epsilon \sqrt{b_1}}<0.
\end{equation}
Clearly, this equation possesses two regimes. The first one,
corresponding to Region I in Ref.~\cite{Martin:2001xs}, is when the
quartic term of the dispersion relation dominates. Expressed in terms
of the new variable $x$, this corresponds to $x^2/4\gg \vert a\vert
$. On the other hand, Region II of Ref.~\cite{Martin:2001xs}
corresponds to a region where the dispersion relation has become
standard and $x^2/4\ll \vert a\vert $. The general solution of
Eq.~(\ref{exact}) (valid in both regions) is given in terms of
parabolic cylinder functions $E(a,x)$ (see Ref.~\cite{gr}) and reads
\begin{equation}
\label{solexact}
\mu = A_1E(a,x)+A_2E^*(a,x),
\end{equation}
where the two constants $A_1$ and $A_2$ are fixed by the initial conditions 
when $\eta \rightarrow -\infty $ (i.e. $x \rightarrow \infty$). In this limit, 
the asymptotic behavior is given by~\cite{gr}
\begin{widetext}
\begin{eqnarray}
\label{limitE}
\lim _{\eta \rightarrow -\infty }\mu &=& 
b_1^{-1/8}\biggl(\frac{\pi }{\epsilon }\biggr)^{-1/4}
\biggl(\frac{n\epsilon }{2\pi }\biggr)^{-1/2}\vert \eta \vert ^{-1/2}
\biggl\{A_1
\exp \biggl[ib_1^{1/2}\frac{n^2\epsilon }{4\pi }\vert \eta \vert ^2
+i\frac{\pi }{4}+i\frac{\phi _2}{2} \biggr]
\nonumber \\
& & +A_2\exp \biggl[-ib_1^{1/2}\frac{n^2\epsilon }{4\pi }\vert \eta \vert ^2
-i\frac{\pi }{4}-i\frac{\phi _2}{2} \biggr]\biggr\},
\end{eqnarray}
where $\phi _2\equiv \arg \Gamma (1/2+ia)$. Thus we 
reproduce the correct WKB behavior if we take
\begin{equation}
\label{A12}
A_1=2^{-1/2}\frac{b_1^{-1/8}}{\sqrt{n}}\biggl(\frac{\pi }{\epsilon }\biggr)^{1/4}
\exp \biggl[-i(b_1)^{1/2}\biggl(\frac{n^2\epsilon }{4\pi }\biggr)
\vert \eta _{\rm i}\vert ^2-i\frac{\pi }{4}-i\frac{\phi _2}{2}\biggr], 
\quad A_2=0.
\end{equation}
\end{widetext}
Let us now study how this solution behaves in Region II. Usually, the
solution in this region is given by $1/\sqrt{2n}e^{\pm in\vert \eta
\vert }$. It is therefore sufficient to have only one branch
proportional to $1/\sqrt{n}$ to recover the standard scale-invariant
spectrum. Using the asymptotic behavior of the parabolic cylinder
functions when $x^2/4 \ll \vert a\vert $, one finds
\begin{equation}
\label{exactII}
\mu \simeq A_1 \biggl(\frac{\pi }{\epsilon }\biggr)^{-1/4}b_1^{1/8}
\exp\biggl[in\vert \eta \vert +i\frac{\pi }{4}\biggr],
\end{equation}
i.e. we precisely recover the conditions necessary to obtain a
Harrison-Zeldovich spectrum, due to the fact that $A_1\propto
1/\sqrt{n}$, see Eq.~(\ref{A12}).
\par

Let us now turn to the case $b_1<0$. An immediate consequence is that
the dispersion relation $n_{\rm eff}(n,\eta )$ vanishes at some point
and then becomes complex. Therefore, we face the following
problems. Firstly, in the region where $n_{\rm eff}(n,\eta )$ is
small, the term $a''/a$ is no longer negligible. In principle it
should be taken into account in the equation of motion but no exact
solution can then be found. However, this problem is not too serious
because the effect of this term on the final spectrum is expected to
be small. Secondly, a much more serious question is the fact that
there is a region where we have to quantize a field in the presence of
imaginary frequency modes. Although imaginary frequencies are standard
in classical physics and in quantum mechanics, they seem to be
problematic in the context of quantum field theory~\cite{Kang}
although there exist concrete physical situations where they are
important~\cite{ssw}. Thirdly, we have to fix the initial conditions
in the complex region. A possible choice for the initial conditions,
which is consistent with the WKB result, is to keep only the
decreasing exponential in the region where the effective frequency
becomes complex as proposed in Ref.~\cite{Martin:2001xs}. Then, an
exact solution in terms of parabolic cylinder functions $U(a,x)$ and
$V(a,x)$ (see Ref.~\cite{gr}) can be found. This gives a
Harrison-Zeldovich spectrum corrected by oscillations and by an
exponential term of the form $e^{An^2}$. This result is in agreement
with what was obtained in Ref.~\cite{Martin:2001xs}. However, this
result rests clearly on ``non-standard physics'' and for this reason
is not so attractive. Therefore, it is important to recall that there
now exist two cases where the final spectrum is modified and
everything is well-defined~\cite{Lemoine:2001ar,BJM}.

\section{Consequences for the matching procedure}

We now investigate what we can learn from the previous exact solution
with respect to the matching between Regions I and II. We concentrate
on the Corley/Jacobson case with $b_1>0$. A priori, two natural
choices for the matching time can be envisaged. The first choice is to
match the solutions when $\lambda =\ell _{_{\rm C}}$. This amounts to
choosing
\begin{equation}
\label{time1}
\vert \eta _{\rm j}\vert =\vert \eta _1 \vert 
=\biggl(\frac{n\ell _{_{\rm C}}}{2\pi \ell _0}\biggl)^{-1}.
\end{equation}
Another possibility is to choose the matching time such that the usual
and the extra contribution in the dispersion relation are equal
\begin{equation}
\label{time1prime}
n^2b_1\biggl(\frac{\ell _{_{\rm C}}}{\lambda }\biggr)^{2}=n^2 
\quad \Rightarrow \quad 
\vert \eta _{\rm j}\vert =\vert \eta _1' \vert =
\biggl(\frac{n\ell _{_{\rm C}}}{2\pi \ell _0}
\biggl)^{-1}b_1^{-1/2}.
\end{equation}
This is in fact equivalent to matching the frequencies $n_{\rm
eff}(n,\eta )$. Therefore, $\eta _1$ and $\eta _1'$ are not equal
unless $b_1=1$. Let us now perform the matching between Regions I and
II. Then, the coefficient $B_1$ is given by:
\begin{widetext}
\begin{eqnarray}
\label{b1}
2inB_1e^{in\eta _{\rm j}} &=& 
2^{-1/2}\biggl( \frac{n\epsilon }{2\pi }\biggr)^{-1/2}
n^{-1/2}b_1^{-1/4}\vert \eta _{\rm j}\vert ^{-1/2} \exp 
\biggl[\mp ib_1^{1/2}\biggl(\frac{n\epsilon }{4\pi \beta }\biggr)n
\biggl(\vert \eta _{\rm j} \vert ^2
-\vert \eta _{\rm i}\vert ^2\biggr)\biggr] 
\nonumber \\
& & \times \biggl \{ in -\frac{1}{2}(1+\beta )\vert \eta _{\rm j}\vert ^{-1}
\pm ib_1^{1/2}\biggl( \frac{n\epsilon }{2\pi }\biggr)n
\vert \eta _{\rm j}\vert \biggr\}.
\end{eqnarray}
For the curly bracket of the above expression one finds
\begin{equation}
\label{bracket}
\{ \dots \}_1 = -\frac{1}{2}(1+\beta )\vert \eta _{\rm j}\vert ^{-1}
+in \biggl(1\pm \biggl \vert \frac{\eta _{\rm j}}{\eta _1'}\biggr \vert 
\biggr).
\end{equation}
In the same manner, one can determine $B_2$ to be
\begin{eqnarray}
\label{b2}
2inB_2e^{-in\eta _{\rm j}} &=& 
2^{-1/2}\biggl( \frac{n\epsilon }{2\pi }\biggr)^{-1/2}
n^{-1/2}b_1^{-1/4}\vert \eta _{\rm j}\vert ^{-1/2} \exp 
\biggl[\mp ib_1^{1/2}\biggl(\frac{n\epsilon }{4\pi \beta }\biggr)n
\biggl(\vert \eta _{\rm j} \vert ^2
-\vert \eta _{\rm i}\vert ^2\biggr)\biggr] 
\nonumber \\
& & \times \biggl \{ in +\frac{1}{2}(1+\beta )\vert \eta _{\rm j}\vert ^{-1}
\mp ib_1^{1/2}\biggl( \frac{n\epsilon }{2\pi }\biggr)n
\vert \eta _{\rm j}\vert \biggr\}.
\end{eqnarray}
\end{widetext}
Again, we can write the curly bracket in the previous expression 
as
\begin{equation}
\label{bracket2}
\{ \dots \}_2 = +\frac{1}{2}(1+\beta )\vert \eta _{\rm j}\vert ^{-1}
+in \biggl(1\mp \biggl \vert \frac{\eta _{\rm j}}{\eta _1'}\biggr \vert 
\biggr).
\end{equation}
The situation is now clear. If the joining is performed at $\eta _{\rm
j}=\eta _1$, the terms proportional to $in$ in the curly brackets have
no reason to cancel out and they are in fact of order one. The term
proportional to $\vert \eta _j\vert ^{-1}$ is very small and can be
neglected. Therefore we reach the conclusion that $B_1\simeq B_2$ and
we have oscillations. However, if we perform the matching at $\eta
_{\rm j}=\eta _1'$ the situation is drastically different. This time
one of the curly brackets vanishes and one of the $B_i$'s becomes of
the order $\eta _1^{'-1}\ll 1$, whereas the other one is of order
$1$. In other words, only one branch survives, we have no oscillations
and the spectrum of fluctuations is unchanged.
\par

A comparison with the exact solution shows that the correct matching
time is $\eta _1'$. Therefore, there are no oscillations in the
spectrum in the case $b_1>0$, as pointed out in
Ref.~\cite{Niemeyer:2001qe}, and the spectrum is unchanged. On the
other hand, for $b_1<0$ the spectrum is modified, in agreement with
the analysis of Ref.~\cite{Martin:2001xs}.

\section{Conclusion}

In this short letter, we have investigated the following technical
points: (i) it has been shown that the method used in
Ref.~\cite{Martin:2001xs} is equivalent to the WKB approach, (ii) a
new solution valid in the case of the Corley/Jacobson dispersion
relation has been presented, (iii) in the case where piecewise
solutions are used, the matching conditions have been studied. It has
been demonstrated that the frequencies rather than the wavelengths
should be matched (as could have been guessed from the equation of
motion) and that if the latter requirement is utilized then artificial
oscillations show up in the spectrum. These results complete the study
of Ref.~\cite{Martin:2001xs}.  
\par

On more general grounds, the conclusion that follows from the previous
considerations is that there is a sensitive dependence of the spectrum
of cosmological fluctuations on the assumptions made at the level of
sub-Planckian physics.  A separate issue is whether these
modifications are reasonable from a physical point of view. This
question cannot be answered in the absence of a realistic theory of
physics beyond the Planck scale. To our knowledge, it is still an open
problem to derive a dispersion relation from, for example, string
theory (see, however, Refs.~\cite{Kempf:2001ac,Amelino-Camelia:2000mn}
for some recent progress).

\medskip
\centerline{\bf Acknowledgements}
\medskip

We are grateful to Steve Corley for stimulating discussions and useful
comments. We acknowledge support from the BROWN-CNRS University Accord
which made possible the visit of J.~M. to Brown during which some of
the work on this project was done, and we are grateful to Herb Fried
for his efforts to secure this Accord.  J.~M. thanks the High Energy
Group of Brown University for warm hospitality. The research was
supported in part by the U.S. Department of Energy under Contract
DE-FG02-91ER40688, TASK A.


\begin{thebibliography}{50}
\bibitem{Martin:2001xs}
J.~Martin and R.~H.~Brandenberger, Phys. Rev. D {\bf 63}, 
123501 (2001), {\tt hep-th/0005209}; R.~H.~Brandenberger and J.~Martin, 
Mod. Phys. Lett. A {\bf 16}, 999 (2001), {\tt astro-ph/0005432}.

\bibitem{Mukhanov:1992me}
V.~F.~Mukhanov, H.~A.~Feldman and R.~H.~Brandenberger, Phys. Rept. 
{\bf 215}, 203 (1992).

\bibitem{U} W.~Unruh, Phys. Rev. D {\bf 51}, 2827 (1995).

\bibitem{Corley:1996ar}
S.~Corley and T.~Jacobson, Phys. Rev. D {\bf 54}, 1568 
(1996), {\tt hep-th/9601073}.

\bibitem{Niemeyer:2001qe}
J.~C.~Niemeyer and R.~Parentani, Phys.Rev. D {\bf 64}, 101301 
(2001), {\tt astro-ph/0101451}.

\bibitem{Niemeyer:2001eh}
J.~C.~Niemeyer, Phys. Rev. D {\bf 63}, 123502 (2001), 
{\tt astro-ph/0005533}.

\bibitem{Starobinsky:2001kn}
A.~A.~Starobinsky, Pisma Zh.\ Eksp.\ Teor.\ Fiz.\  {\bf 73}, 415 (2001)
[JETP Lett.\  {\bf 73}, 371 (2001)], {\tt astro-ph/0104043}.

\bibitem{Martin:2000bv}
J.~Martin and R.~H.~Brandenberger, {\tt astro-ph/0012031}.

\bibitem{Kowalski-Glikman:2001dz}
J.~Kowalski-Glikman, Phys. Lett. B {\bf 499}, 1 (2001), {\tt astro-ph/0006250}.

\bibitem{Mersini:2001su}
L.~Mersini, M.~Bastero-Gil and P.~Kanti, Phys. Rev. D {\bf 64}, 043508 
(2001), {\tt hep-ph/0101210}.

\bibitem{Alexander:2001ck}
S.~Alexander and J.~Magueijo, {\tt hep-th/0104093}.

\bibitem{Alexander:2001dr}
S.~Alexander, R.~Brandenberger and J.~Magueijo, {\tt hep-th/0108190}.

\bibitem{Lemoine:2001ar}
M.~Lemoine, M.~Lubo, J.~Martin and J.~Uzan, Phys. Rev. D {\bf 64}, 
043508 (2001), {\tt hep-th/0109128}.

\bibitem{Kempf:2001ac}
A.~Kempf, Phys. Rev. D {\bf 63}, 083514 (2001), {\tt astro-ph/0009209}.

\bibitem{Chu:2000ww}
C.~Chu, B.~R.~Greene and G.~Shiu, Mod. Phys. Lett. A {\bf 16}, 2231 (2001), 
{\tt hep-th/0011241}.

\bibitem{Kempf:2001fa}
A.~Kempf and J.~C.~Niemeyer, Phys. Rev. D {\bf 64} 103501 (2001), {\tt 
astro-ph/0103225}.

\bibitem{Easther:2001fi}
R.~Easther, B.~R.~Greene, W.~H.~Kinney and G.~Shiu, Phys. Rev. D {\bf 64} 
103502 (2001), {\tt hep-th/0104102}.

\bibitem{BJM} R.~H.~Brandenberger, S.~E.~Joras and J.~Martin, 
{\tt hep-th/0112122}.

\bibitem{Tanaka:2000jw}
T.~Tanaka, {\tt astro-ph/0012431}.

\bibitem{ABM} L.~R.~W. Abramo, R.~H.~Brandenberger and V.~Mukhanov, 
Phys. Rev. D {\bf 56}, 3248 (1997); V.~Mukhanov, L.~R.~W.~Abramo
and R.~H.~Brandenberger, Phys. Rev. Lett. {\bf 78}, 1624 (1997).

\bibitem{gr} I.~S.~Gradshteyn, I.~M.~Ryzhik, ``Table of Integrals, Series, and
Products'', Sixth edition (2000), Academic Press.

\bibitem{Kang} G.~Gang, {\tt hep-ph/9603166}.

\bibitem{ssw} L.~I.~Schiff, H.~Snyder and J.~Weinberg, Phys. Rev. {\bf 57} 
315 (1940).

\bibitem{Amelino-Camelia:2000mn} G.~Amelino-Camelia, {\tt gr-qc/0012051}.





\end{thebibliography}
\end{document}